\documentclass[mathleft
]{an}
\usepackage{graphicx}
\usepackage{times}
\def\gtsima{$\; \buildrel > \over \sim \;$}
\def\ltsima{$\; \buildrel < \over \sim \;$}
\def\gtrsim{\lower.5ex\hbox{\gtsima}}
\def\lesssim{\lower.5ex\hbox{\ltsima}}

\begin{document}

\Pagespan{1}{}
\Yearpublication{????}%
\Yearsubmission{2010}%
\Month{??}%
\Volume{???}%
\Issue{??}%

\title{The metallicity of the nebula surrounding the ultra-luminous
X-ray source NGC~1313~X-2}

\author{E.~Ripamonti
\inst{1}\fnmsep\thanks{Corresponding authors:
\email{ripamonti.e@gmail.com}\newline}
\and  M.~Mapelli\inst{1}
\and  L.~Zampieri\inst{2}
\and  M.~Colpi\inst{1}
}
\titlerunning{Metal abundances around NGC~1313~X-2}
\authorrunning{Ripamonti et al.}
\institute{
Dipartimento di Fisica `G. Occhialini', Universit\`a di Milano-Bicocca, Piazza della Scienza 3, I--20126 Milano, Italy 
\and 
INAF-Osservatorio astronomico di Padova, Vicolo dell'Osservatorio 5, I--35122, Padova, Italy}

\received{31 August 2010}
\accepted{???}
\publonline{later}


\abstract{%
Recent models of the formation of ultra-luminous X-ray sources (ULXs)
predict that they preferentially form in low-metallicity
environments. We look at the metallicity of the nebula
surrounding NGC 1313~X-2, one of the best-studied ULXs. Simple
estimates, based on the extrapolation of the metallicity gradient
within NGC 1313, or on empirical calibrations (relating metallicity to
strong oxygen lines) suggest a quite low metal content
($Z \sim 0.1\, Z_\odot$). But such estimates do not account for
the remarkably strong X-ray flux irradiating the nebula. Then, we
build photoionization models of the nebula using CLOUDY; using such
models, the constraints on the metallicity weaken substantially, as we
find $0.15\, Z_\odot \le Z \le 0.5\, Z_\odot$.}

\maketitle

\section{Introduction}
The number of ultra-luminous X-ray sources (ULXs) in a galaxy appears
to correlate with its star formation rate (SFR; see e.g. Irwin,
Bregman \& Athey 2004). Such correlation is reminiscent of the one
between high-mass X-ray binaries (HMXBs) and SFR (e.g. Grimm, Gilfanov
\& Sunyaev 2003; Ranalli, Comastri \& Setti 2003; Mineo, Gilfanov \&
Sunyaev 2010).  On the other hand, in the last decade it has been
suggested that ULXs correlate with low metallicities (Pakull \&
Mirioni 2002; Zampieri et al. 2004; Soria et al. 2005; Swartz, Soria
\& Tennant 2008). In fact, low metallicity would affect stellar
evolution: it might increase the maximum stellar remnant mass
(Mapelli, Colpi \& Zampieri 2009; Zampieri \& Roberts 2009; Mapelli et
al. 2010a, hereafter M10), or increase the number and luminosity of
HMXBs in more subtle ways (Linden et al. 2010). Indeed, M10 looked at
the observational properties (number of ULXs, SFR, and metallicity
$Z$) of a sample of galaxies, and found that metal-poor ($Z\lesssim
0.2Z_\odot$) galaxies appear to host more ULXs than metal-rich
galaxies with similar SFR (see also Mapelli et al. 2010b), even if
the statistical significance of this result is low.

It must be remarked that studies such as M10 necessarily rely upon
some {\it average} metallicity of each X-ray-observed galaxy, rather
than on the metallicity of the ULXs themselves: otherwise, the sample
would be far too small for any statistical analysis. However,
metallicity variations within the same galaxy are well known: almost
all the large spiral galaxies in our vicinity exhibit a metallicity
gradient from the centre to the periphery (e.g. Pilyugin, V\'ilchez \&
Contini 2004); furthermore, even in smaller objects, or at fixed
distances from the galactic centre, the spread in metallicities is
often larger than the $\sim 0.1$ dex uncertainty typically associated
with metallicity measurements. Then, it is important to determine the
metallicity of ULXs (or at least of their immediate environments),
since this would provide a more direct test of some of the models
above (e.g., the model presented by M10 suggests that it is very
difficult to form ULXs in environments with $Z\gtrsim 0.4\,Z_\odot$).

Unfortunately, the number of ULXs with known optical counterparts is
very small; furthermore, spectroscopic observations of ULXs and/or
their environments are available only for a handful of objects (NGC
1313 X-2, IC 342 X-1, Holmberg II X-1, among others; see
e.g. Mucciarelli et al. 2005, Lehmann et al. 2005, Gris\`e, Pakull \&
Motch 2006, Abolmasov et al. 2007, Feng \& Kaaret 2008), and often do
not have a sufficient spectral coverage for a reliable metallicity
estimate. In practice, the only object where such determination is
possible with data in the literature is NGC 1313 X-2. Here we will aim
to directly estimate the metallicity of the gaseous nebula surrounding
such object.

\section{The environment of NGC1313 X-2}
NGC 1313 is a barred spiral at a distance of 3.7 Mpc (Tully 1988),
with moderate SFR ($\simeq 1.4\, {\rm M_\odot\, yr^{-1}}$, as inferred
from the H$\alpha$ measurements of Ryder \& Dopita 1994, using the
Kennicutt 1998 conversion). It hosts 3 ULXs (Colbert et al. 1995), one
of whom is a known supernova remnant. Observations by Pagel, Edmunds
\& Smith (1980), and by Hadfield \& Crowther (2007) allow to determine
the oxygen abundances\footnote{The quantity 12+log(O/H) is generally
  used as a proxy for total metallicity; in this notation a
  metallicity $Z=Z_\odot \equiv 0.02$ corresponds to
  12+log(O/H)=8.92.} of 8 HII regions with the reliable $T_e$ method
(e.g . Pilyugin 2001; Pilyugin \& Thuan 2005 - herafter PT05).  A
linear fit to these metallicity data leads to a gradient
\begin{equation}
12+\log({\rm O/H}) = (8.52\pm0.13) - (0.55\pm0.25)(r/r_{{\rm 25}}),
\label{eq_gradient}
\end{equation}
where $r$ is the elliptical radius, and $r_{25} \simeq 4.56'$ is the
optical semi-major axis of the galaxy (de Vaucouleurs et al. 1991);
this corresponds to $Z\sim0.4\,Z_\odot$ at the centre, and $Z\sim
0.16\,Z_\odot$ at the location of the outermost HII region used for
the fit ($r \simeq 0.75 r_{\rm 25}$).

The source X-ray luminosity (0.3--10 keV) is known to vary in the
2--30 $\times 10^{39}\, {\rm erg\, s^{-1}}$ range. The X-ray emission
is generally fit with a multicolor disk (MCD, with $kT_{\rm MCD}$
varying in the 0.13--0.25 keV range) plus power law (PL, with photon
index varying between 1.7 and 2.5) spectrum, with the energy in the
MCD component representing a fraction 0.46--0.94 of the energy in the
PL component; such spectrum is then absorbed by a column density
$N_{\rm H,t}$ in the 2--4$\times 10^{21}\,{\rm cm^{-2}}$ range
(e.g. Zampieri et al. 2004, Feng \& Kaaret 2006, Mucciarelli et
al. 2007, and references therein), much larger than the galactic
absorption towards NGC~1313 ($N_{\rm H,g}\simeq 4\times10^{20}\,{\rm
  cm^{-2}}$, Dickey \& Lockman 1990).

The optical counterpart to NGC1313 X-2 has been identified as object
C1 (Pakull, Gris\'e \& Motch 2006, Mucciarelli et al. 2007, Liu et
al. 2007), using HST and VLT observations. Assuming E(B-V)$\sim$0.1
(Mucciarelli et al. 2007, Gris\'e et al. 2008)\footnote{Liu et
al. (2007) suggested E(B-V)$\sim$0.3, but according to Patruno \&
Zampieri (2010) such value is incompatible with other observed
properties.}, the object absolute V magnitude is $M_{\rm V}\simeq
-4.6\pm0.2$, and its blue color ($B-V=-0.13\pm0.06$) suggests an high
effective temperature (20000 K$\le T_{\rm eff} \le$ 30000 K). This is
surrounded by an OB association (about 20 Myr old), and by a large
ionized nebula (about $500\times300$ pc in size; cfr. Gris\'e et
al. 2008).  Remarkably, the optical luminosity of object C1 appears
to vary with a periodicity of $\sim$ 6.1 days (Liu, Bregman \&
McClintock 2009). Such result is still uncertain (e.g. Impiombato et
al. 2010), but suggests that the accreting object is a black hole with
$M\sim$ 50--100$\,{\rm M_\odot}$ (Patruno \& Zampieri 2010).

\section{Simple metallicity estimates}
NGC 1313 X-2 is located about 6 arcmin south of the centre of NGC1313,
corresponding to $r \simeq 1.5 r_{\rm 25}$: eq. (\ref{eq_gradient})
would imply 12+log(O/H)=$7.69\pm0.40$, i.e $Z\simeq
0.06^{+0.09}_{-0.04}\,Z_\odot$.

In order to test this result, we looked at the VLT spectrum of the
optical counterpart of NGC1313 X-2 presented in Mucciarelli et
al. (2005). In order to extract the nebular continuum, we looked
specifically at those positions along the slit which are close to NGC
1313 X-2, but where no stellar emission can be detected. In Table
\ref{tab_lines} we list the emission lines we were able to confidently
detect; the typical detection limit was at about $0.2 I({\rm H}\beta)$.
It should be noted that several emission lines exhibit
significant variations along the slit, so that the intensity ratios in
Table \ref{tab_lines} are the result of spatial averaging.

\begin{table}
\begin{center}
\caption{List of detected emission lines.}
\label{tab_lines}
\begin{tabular}{lcc}\hline
Ion            & $\lambda ({\rm \AA})^{\rm a}$ & $I({\rm line})/I({\rm H}\beta)^{\rm b}$\\ 
\hline
OII            & 3727            & $6.03\pm1.34$\\
HI (H$\delta$) & 4102            & $0.38\pm0.16$\\
HI (H$\gamma$) & 4340            & $0.56\pm0.16$\\
HI (H$\beta$)  & 4861            & $1.00\pm0.17^{\rm c}$\\
OIII           & 4959            & $0.47\pm0.15$\\
OIII           & 5007            & $1.69\pm0.36$\\
OI             & 6300            & $0.93^{+0.23}_{-0.46}$ $^{\rm d}$\\
NII            & 6548            & $0.12\pm0.10^{\rm d}$\\
HI (H$\alpha$) & 6563            & $3.10\pm0.70^{\rm e}$\\
NII            & 6584            & $0.53\pm0.30^{\rm d}$\\
SII            & 6725            & $1.95\pm0.42$\\
\hline
\end{tabular}
\end{center}
\footnotesize{$^{\rm a}$ Rest-frame wavelength. $^{\rm b}$ Ratio of
  de-reddened fluxes; the de-reddening was performed by imposing that
  I(H$\alpha$)/I(H$\beta$)=3.1 (cfr. Osterbrock 1989). Quoted errors
  include uncertainties on the line itself, on the normalization -
  i.e. on I(H$\beta$) - and on the reddening, unless differently
  specified. $^{\rm c}$ Error bars include only the uncertainty on the
  line itself. $^{\rm d}$ Error bars include an additional component
  due to the removal of the blending of lines. $^{\rm e}$ Error bars
  include uncertainties in the line itself, in the normalization to
  I(H$\beta$), and in the de-blending.}
\end{table}

Since we were unable to observe the weak OIII $\lambda=4363\,{\rm
  \AA}$ line, the simplest way to estimate the metallicity of the
nebula is to use the P and ff methods described by PT05, relying only
upon the intensities of strong oxygen lines (i.e. the indexes
$R_2 \equiv I({\rm OII}3727)/I({\rm H}\beta)\simeq 6.03$,
$R_3 \equiv [I({\rm OIII}4959)+I({\rm OIII}5007)]/I({\rm H}\beta) \simeq
2.16$, their sum $R_{23} \equiv R_2+R_3\simeq 8.19$, and the ratio
$P \equiv R_3/R_{23}\simeq 0.26$).
In this case, the $P$ method fails, since the object falls in the
``gap'' between the high-Z and the low-Z branches of the P calibration
(a fact weakly suggesting that $8.0\lesssim 12+\log({\rm O/H})
\lesssim 8.25$, i.e. $0.12 Z_\odot \lesssim Z \lesssim 0.21 Z_\odot$);
instead, the ff method gives $12+\log({\rm O/H})\simeq 7.90$, or
$Z\simeq 0.10 Z_\odot$.

However, such metallicity is very uncertain, not only because the
results of the P and the ff methods are (slightly) inconsistent, but
especially because both methods were calibrated with {\it normal} HII
regions, where ionization is (almost) completely due to UV photons
emitted by O-B stars. Instead, in this case we know that the nebula is
in the immediate vicinity of a very intense X-ray source, so that a
significant fraction of the ionization might be collisional, as it is
induced by much harder photons (in fact, the bulk of the ionizations
are due to the collisions involving the energetic photo-electrons
produced by the X-ray photons). Furthermore, the edge of normal HII
regions is much sharper than that of X-ray ionized nebulae, since
X-ray photons can travel much farther than UV photons in a neutral
gas: as a result, there is a large volume with moderate (0.1--0.9)
ionization fraction, producing unusually strong OI $\lambda=6300\,{\rm
  \AA}$ emission. Such line is indeed observed, even if shocks might
provide a further (or alternative) explanation for its strength.

\section{Photoionization models}
Since ``standard'' line strength-metallicity calibrations (such as the
ones in PT05) are likely inappropriate for this object, we used the
public code CLOUDY (version 08.00; see Ferland et al. 1998) in order
to build a grid of photoionization models of the nebula, and then
compare the prediction of these models to the observed
spectrum\footnote{We note that (differently from the MAPPINGS code -
  see e.g. Sutherland \& Dopita 1993, and Allen et al. 2008), CLOUDY
  does not account for shocks. This might affect our results, so that
  we plan to extend this work by using MAPPINGS in the near future.}.

\begin{table}
\begin{center}
\caption{Parameter space explored by the CLOUDY models.}
\label{tab_models}
\begin{tabular}{ll}\hline
Parameter        & List of values\\
\hline
$T_*$ [K]                              & 26200, 30000, 31000, 32000,\\
                                       & 33500, 35000, 40000, 45000\\
$N_{\rm H,i}\ {\rm [10^{21}\,cm^{-2}]}$  & 0.01, 0.03, 0.1, 0.3, 1., 2.\\
12+log(O/H)                       & 9.22, 8.92, 8.82, 8.72, 8.62, 8.52, 8.42,\\
                                       & 8.32, 8.22, 8.12, 7.92, 7.62, 7.32\\
$[$N/O$]$                              & 0, -0.5\\
$[$dust/metal$]$                       & 0, -1\\
$\log(n_{\rm H})\ {\rm [cm^{-3}]}$      & 0, 0.5, 1, 1.5, 2\\
$L_{\rm X}\ {\rm [erg\,s^{-1}]}$        & $6\times10^{39}$\\
$kT_{\rm MCD}$ [keV]                    & 0.2\\
$\Gamma_{\rm PL}$                       & 2.0\\
$F_{\rm MCD}/F_{\rm PL}$                 & 0.8\\
$N_{\rm H,t}\ {\rm [10^{21}\, cm^{-2}]}$ & 3.0\\
$N_{\rm H,g}\ {\rm [10^{21}\, cm^{-2}]}$ & 0.4\\
$f_{\rm f}$                             & 1/3\\
$R_{\rm in}$ [pc]                       & 1\\          
\hline
\end{tabular}
\end{center}
\footnotesize{$T_{*}$ is the effective temperature of the companion
  star; its ionizing flux was varied accordingly, as in Tab. 2.3 of
  Osterbrock (1989). $N_{\rm H,i}$ is the column density which is
  crossed by the radiation before entering the nebula (i.e., within
  the source or in its immediate vicinity). 12+log(O/H) is the nebula
  oxygen abundance, and indicates the general metallicity, as we
  assume the ratio X/O of any metal X (except N) to remain equal to
  the solar value. [N/O]$\equiv\log[({\rm N/O})/({\rm N/O})_\odot]$, and
  [dust/metal] is defined in the same way, but with reference to the
  dust/metal ratio in Orion, rather than in the Sun. $n_{\rm H}$ is
  the hydrogen number density in the nebula. $L_{\rm X}$ is the
  unabsorbed X-ray luminosity in the 0.3--10 keV spectral
  range. $kT_{\rm MCD}$, $\Gamma_{\rm PL}$, and $F_{\rm MCD}/F_{\rm
    PL}$ are the temperature of the multi-color disk, the power-law
  photon index, and the ratio of the energy emitted in the two
  components, respectively. $N_{\rm H,t}$ and $N_{\rm H,g}$ are
  the total column density from the X-ray source to the observer, and
  the column density due to absorption in our Galaxy,
  respectively. Finally, $f_{\rm f}$ is the filling factor, and
  $R_{\rm in}$ is the distance of the inner ``edge'' of the nebula
  from the X-ray source.}
\end{table}

Table \ref{tab_models} illustrates the region of parameter space
covered by the $\sim$ 12000 photoionization models of the grid. We
explored variations in the spectrum and luminosity of the companion
star\footnote{Each of the values of the effective stellar temperature
  $T_{*}$ is paired to the UV luminosity of a star of the
  corresponding spectral type; spectra were based on stellar
  atmospheres from a library built with the ATLAS9 code (Castelli \&
  Kurucz 2004) where we assumed that the metallicity of the companion
  star is equal to that of the nebula.}, in the intrinsic absorption
of the X-ray spectrum (which was assumed to be negligible at photon
energies below $54.4\,{\rm eV}$), in the metallicity of the gas
(mainly expressed by the oxygen abundance), in the N/O and dust/metal
ratios, and in the density of the nebula. Other quantities (such as
the X-ray luminosity $L_{\rm X}$) were kept constant at their observed
values (or, when they are known to vary, at reasonable averages of the
observed values), or at ``typical'' values observed in other HII
region (e.g. the filling factor). Finally, we assumed the nebula to be
a spherical corona centered on the X-ray source, with an arbitrary
inner radius $R_{\rm in}$, and an outer radius which is set by the
relation
\begin{equation}
R_{\rm out} = R_{\rm in} +
{{N_{\rm H,t} - N_{\rm H,i} - N_{\rm H,g}}\over
{n_{\rm H}\; f_{\rm f}\; (Z/Z_\odot)}}.
\end{equation}

By comparing the model predictions with the observations (we actually
ranked our models, with ``scores'' assigned with a procedure similar
to the one used to estimate the $\chi^2$ statistics\footnote{However,
  such ``scores'' do not enjoy the properties of $\chi^2$
  statistics.}), we found that a very good agreement could be obtained
with model A (cfr. Table \ref{tab_modelcomparison}). In such model,
$T_{*}=40000\,{\rm K}$, $N_{\rm H,i}=10^{20}\,{\rm cm^{-2}}$,
$12+\log({\rm O/H})=8.22$ (i.e. $Z=0.2\,Z_\odot$), [N/O]=-0.5,
[dust/metal]=-1, $n_{\rm H}=1\, {\rm cm^{-3}}$.  Apart from moderate
($\sim1\sigma$) discrepancies for the OIII and SII lines, the main
problem of this model is an over-prediction of the radius of the HII
region ($\sim 400\,{\rm pc}$); but the overall agreement is quite
good.

However, there is an ample range of models with good overall
agreement. For example, in Table \ref{tab_modelcomparison} we show the
result for a second model, quite different from model A. In fact,
model B was calculated with $T_{*}=35000\,{\rm K}$, $N_{\rm H,i} =
3\times10^{20}\,{\rm cm^{-2}}$, $12+\log({\rm O/H})=8.92$ (i.e. $Z=Z_\odot$),
[N/O]=-0.5, [dust/metal]=-1, $n_{\rm H}=1\,{\rm cm^{-3}}$. There is
excellent agreement in all the lines except the NII ones, which are
significantly ($\ge2\sigma$) over-predicted (instead, the radius of
the HII region is predicted to be $\sim 150\,{\rm pc}$).

\begin{table}
\begin{center}
\caption{Comparison of observations and models.}
\label{tab_modelcomparison}
\begin{tabular}{cccccc}
\hline
Model & OII           & OIII          & OI            & NII           & SII\\
\hline
A     & 5.76          & 2.60          & 0.87          & 0.59          & 1.45\\
      & $-0.20\sigma$ & $+0.99\sigma$ & $-0.12\sigma$ & $-0.21\sigma$ & $-1.19\sigma$\\
B     & 5.77          & 2.04          & 0.60          & 1.27          & 1.90\\
      & $-0.20\sigma$ & $-0.27\sigma$ & $-0.71\sigma$ & $+2.39\sigma$ & $-0.12\sigma$\\
\hline
\end{tabular}
\end{center}
\footnotesize{For each model, in the first line we report the sum of
  the predicted intensities (normalized to H$\beta$) of the lines of
  each ion which are listed in Table 1; in the second line we report
  the difference between observed values and model predictions, in
  terms of the observational uncertainty ($\sigma$).}
\end{table}

When we marginalize over all the free parameters, we infer that both
[N/O] and [dust/metal] must be significantly smaller than 0 (no model
with [N/O]=0 can be considered satisfactory; the effects of
[dust/metal]=0 are less dramatic, but in {\it all} the comparisons of
pairs of models differing only for this quantity, the low-dust one is
better than the high-dust case). With similar arguments, we find that
$35000 \lesssim T_{*}/{\rm K} \lesssim 40000$, $0.1 \lesssim N_{\rm
  H,i}/(10^{21}\,{\rm cm^{-2}}) \lesssim 0.3$, and $1 \lesssim
n_{\rm H}/({\rm cm^{-3}}) \lesssim 10$. Unfortunately, the less
constraining result is the one about the metallicity, as we find that
there exist satisfactory models for $8.12\lesssim 12+\log({\rm O/H})
\lesssim 8.92$ (i.e. for $0.15 \le Z/Z_\odot \le 1$).

The metallicity result can be improved by remarking that theoretical
considerations on the production of N (e.g. Matteucci 1986) suggest
that, while some N is of ``primary'' origin (i.e. it can be produced
directly from metal-free stars), it is also a ``secondary'' element
(i.e., its production requires the presence of other metals). Such
fact implies that (N/O) should positively correlate with (O/H), as is
actually observed (e.g. Pilyugin, Thuan \& V\'ilchez 2003). Indeed, if
we look at fig. 3 of Pilyugin et al. 2003, we see that no HII region
has been observed with both 12+log(O/H)$\gtrsim$8.62
(i.e. $Z\gtrsim0.5Z_\odot$) and [N/O]$\lesssim$-0.5. Then, the
``satisfactory'' models with $Z \gtrsim 0.5Z_\odot$ are likely
unphysical, since all of them have [N/O]=-0.5.

\section{Summary and conclusions}
We investigated the metal abundance in the nebula surrounding the ULX
source NGC 1313 X-2 by building detailed photoionization models with
CLOUDY, and comparing their predictions to observational data. The
resulting metallicity, $Z\sim 0.2^{+0.3}_{-0.05}\,Z_\odot$, and
especially its upper limit, is quite higher than what could be
inferred from simpler methods (both the metallicity gradient in
NGC~1313, and the simple metallicity calibrations suggest $Z \sim 0.1
Z_\odot$). This might partially derive from the known discrepancy
between ``empirical'' and ``model'' abundances (see e.g. the
discussion at the end of Moustakas et al. 2010), but more likely
reflects the inadequacy of empirical calibrations in environments
where the ionization is due to the radiation emitted by an X-ray
source, rather than normal stars.

Other possible sources of error are i) the possible presence of shocks
(Pakull et al. 2010, and Russel et al. 2010 showed that shocks exist
at least in some of the nebulae surrounding ULXs), and ii) the fact
that the optical-UV spectrum of the counterpart of NGC~1313~X-2 is
likely different from that of a normal star, because it definitely
includes a component due to the reprocessing of the X-ray radiation
(e.g. Patruno \& Zampieri 2008,2010). In the near future we plan to
extend our analysis, including both these effects.

\acknowledgements We thank A.~Bressan, P.~Marigo, the organizers and
the participants to the conference ``Ultra-Luminous X-ray sources and
Middle Weight Black Holes'' (Madrid, 24th-26th May 2010) for useful
discussions. LZ and MC acknowledge financial support through INAF
grant PRIN-2007-26".

\end{document}